\documentclass[sn-aps]{sn-jnl}
\usepackage{xcolor}
\usepackage[capitalise]{cleveref}
\usepackage{dsfont}
\usepackage{bm}
\usepackage{physics}
\usepackage{enumerate}

\renewcommand{\l}{\ell}
\newcommand{\pon}{k}
\newcommand{\chiEFT}{$\chi$EFT}
\newcommand{\mN}{m_N}
\newcommand*{\LO}{LO}
\newcommand*{\NLO}{NLO}
\newcommand*{\NNLO}{N$^2$LO}
\newcommand*{\NNNLO}{N$^3$LO}

\newcommand*{\TLO}{T^{(0)}}
\newcommand*{\TNLO}{T^{(1)}}
\newcommand*{\TNNLO}{T^{(2)}}
\newcommand*{\TNNNLO}{T^{(3)}}

\newcommand*{\VLO}{V^{(0)}}
\newcommand*{\VNLO}{V^{(1)}}
\newcommand*{\VNNLO}{V^{(2)}}
\newcommand*{\VNNNLO}{V^{(3)}}

\newcommand*{\tS}{^3S_1 \mathrm{-}^3D_1}
\newcommand*{\np}{$np$}
\newcommand*{\Tl}{T_\mathrm{lab}}
\newcommand*{\NN}{$NN$}
\newcommand*{\dtS}{\delta_{3S1}}

\jyear{2024}%

\begin{document}

%\title[ ]{Low-energy theorems for effective range parameters in $\chi$EFT using a perturbative power counting}

\title[ ]{Low-energy theorems for neutron-proton scattering in $\chi$EFT using a perturbative power counting}
%\title[]{Low-energy theorems in chiral effective field theory using a perturbative power counting}

%%=============================================================%%
%% Prefix	-> \pfx{Dr}
%% GivenName	-> \fnm{Joergen W.}
%% Particle	-> \spfx{van der} -> surname prefix
%% FamilyName	-> \sur{Ploeg}
%% Suffix	-> \sfx{IV}
%% NatureName	-> \tanm{Poet Laureate} -> Title after name
%% Degrees	-> \dgr{MSc, PhD}
%% \author*[1,2]{\pfx{Dr} \fnm{Joergen W.} \spfx{van der} \sur{Ploeg} \sfx{IV} \tanm{Poet Laureate} 
%%                 \dgr{MSc, PhD}}\email{iauthor@gmail.com}
%%=============================================================%%

\author*[1]{\fnm{Oliver} \sur{Thim}\email{oliver.thim@chalmers.se}}

\affil*[1]{\orgdiv{Department of Physics}, \orgname{Chalmers University of Technology}, \orgaddress{\city{Göteborg}, \postcode{SE-412 96}, \country{Sweden}}}

\abstract{Low-energy theorems (LETs) for effective-range parameters in nucleon-nucleon scattering encode properties of the long-range part of the nuclear force. We compute LETs for \textit{S}-wave neutron-proton scattering using chiral effective field theory with a modified version of Weinberg power counting. Corrections to the leading order amplitude are included in distorted-wave perturbation theory and we incorporate contributions up to the third order in the power counting. We find that LETs in the $^1S_0$ and $^3S_1$ partial waves agree well with empirical effective-range parameters. At the same time, phase shifts up to laboratory scattering energies of about 100~MeV can be reproduced. We show that it is important to consider the pion mass splitting in the one-pion exchange potential in the $^1S_0$ partial wave while the effect is negligible in the $^3S_1$ partial wave. We conclude that pion exchanges, as treated in this power counting, accurately describe the long-range part of the $S$-wave nuclear interaction.
}


\maketitle
\newpage
\section{Introduction}\label{sec:introduction}
The importance of the low-energy limit in quantum mechanical scattering dates back to the work of H. Bethe \cite{Bethe:1949yr}, who realized that two-particle scattering at sufficiently low energies possesses a universal behavior characterized by only two parameters: the scattering length and effective range. This universal theory was extended to what is known as effective range theory by utilizing analytical properties of the scattering amplitude. Besides successfully parameterizing low-energy scattering amplitudes, effective range theory can serve as a tool to analyze the low-energy properties of nuclear interaction models derived from effective field theories (EFTs) \cite{Cohen:1998jr}.

In seminal works \cite{Weinberg:1990rz,Weinberg:1991um}, Weinberg  proposed to apply an EFT description to model the nuclear force based on the spontaneously broken chiral symmetry of low-energy Quantum Chromodynamics (QCD). In this EFT, pion-nucleon interactions are governed by a Lagrangian consistent with the low-energy symmetries of QCD. Nuclear interaction potentials can be constructed by assessing the relative importance of the emerging effective interactions in what is known as power counting (PC) \cite{Epelbaum:2008ga,Machleidt:2011zz,Hammer:2019poc}. The PC is performed in terms of $(Q/\Lambda_b)^\nu$, where $Q$ denotes the relevant low-energy scales and $\Lambda_b$ the breakdown scale of the EFT. In nucleon-nucleon (\NN) scattering $Q\sim \{p, m_\pi\}$, where $p$ is the modulus of the \NN{} relative momentum and $m_\pi$ the average pion mass. We refer to $\nu$ as the chiral order, and specifically $\nu=0$ as leading order (LO), $\nu=1$ as next-to-leading order (NLO), and so on. 
The resulting EFT, known as chiral effective field theory (\chiEFT), has been employed in the last decades to develop two- and three-nucleon potentials to high chiral orders \cite{Ordonez:1995rz,vanKolck:1994yi,Ordonez:1993tn,Entem:2015xwa,Reinert:2017usi,Epelbaum:2002vt}. Low-energy constants (LECs) parametrize the unknown coupling strengths in the Lagrangian, and thus appear in the potentials, and need to be inferred from data. The \chiEFT{} potentials in combination with computational advancements in solving the many-body Schrödinger equation have enabled quantitative EFT predictions of nuclear properties across the nuclear chart, see e.g. Refs~\cite{Hagen:2015yea,Arthuis:2020toz,Hu:2021trw}. 

To date, Weinberg PC (WPC) is the main PC being used to construct quantitative chiral potentials. However, an ongoing effort to construct alternative PCs has been pursued to construct renormalization-group (RG) invariant interactions. The amplitudes in WPC are not RG-invariant in the sense that the momentum space cutoff used to regulate the divergences cannot be taken arbitrarily large. This can be traced to the non-perturbative treatment of singular potentials, which leads to uncontrolled divergences if sufficient counterterms are not included \cite{Nogga:2005hy,vanKolck:2020llt,Long:2007vp,PhysRevA.64.042103} (see, e.g., Ref.~\cite{Epelbaum:2018zli} for another approach to deal with the singular interactions in \chiEFT). In the late 1990s, Kaplan et al.~\cite{Kaplan:1998tg,Kaplan:1998we} proposed an RG-invariant PC where pions are treated perturbatively in all \NN{} partial waves. Despite some success in describing scattering phase shifts, further studies showed that the convergence radius in terms of scattering energy was not improved compared to a theory without pions \cite{Fleming:1999ee}. 

Cohen and Hansen \cite{Cohen:1998jr,Cohen:1999iaa} further investigated this PC (often referred to as KSW counting) by studying the $S$-wave effective range expansion (ERE) for the neutron-proton (\np{}) scattering amplitude. They showed that the  higher-order ERE parameters beyond the scattering length and effective range are predicted functions of the scattering length, nucleon mass, and the parameters defining the long-range pion-exchange potential. By calibrating the unknown LECs to reproduce the empirical scattering length and effective range, the higher-order ERE parameters are thus predictions solely governed by the long-range part of the interaction. We will refer to such predictions of ERE parameters as low-energy theorems (LETs) \cite{Cohen:1998jr,Cohen:1999iaa}. Cohen and Hansen compared the LETs with empirical ERE parameters extracted from the Nijmegen partial wave analysis \cite{Stoks:1993tb}, where the latter provides a robust parametrization of the low-energy behavior of the nuclear force \cite{NavarroPerez:2014ovp}. The comparison showed a poor agreement between the LETs and the empirical ERE parameters while the PC produces realistic phase shifts \cite{Kaplan:1998tg,Kaplan:1998we}. 
The main conclusion of Refs.~\cite{Cohen:1998jr,Cohen:1999iaa} was that LETs serve as a non-trivial test to identify if the long-range part of the potential induces a correct near-threshold energy dependence of the scattering amplitude. 

Following Ref.~\cite{Cohen:1998jr}, LETs have been used as a tool to study the low-energy properties of EFT potentials \cite{Ando:2011aa,PavonValderrama:2003np,Epelbaum:2004fk} and Ref.~\cite{Epelbaum:2003xx} showed that LETs computed in WPC has a good agreement with empirical ERE parameters extracted in Ref.~\cite{Stoks:1993tb}. Furthermore, in \cite{Epelbaum:2009sd} LETs are used as a tool to study renormalization problems; and Refs.~\cite{Baru:2015ira,Baru:2016evv} derive low-energy theorems for a varying pion mass and apply them to analyze lattice-QCD calculations. 

In this paper, we study LETs for \np{} scattering in $S$-waves up to next-to-next-to-next-to-leading order (\NNNLO)\footnote{Note that the $\nu=1$ contribution does not vanish in this PC as opposed to WPC. Hence, NLO and \NNLO{} in WPC corresponds to \NNLO{} and \NNNLO{} in the Long and Yang PC.} applying the perturbative and RG-invariant modified WPC (MWPC) proposed by Long and Yang~\cite{Long:2012ve,PhysRevC.84.057001,PhysRevC.85.034002}. It has been demonstrated that phase shifts, \np{} scattering observables and binding energies in $A=3,4$ nuclei can be described in the Long and Yang PC \cite{Long:2012ve,PhysRevC.84.057001,PhysRevC.85.034002,Thim:2023fnl,Thim:2024yks,Yang:2020pgi}, but LETs have not been investigated. We want to study LETs to quantify the accuracy to which pion exchanges describe the low-energy part of the two-nucleon interaction in this PC. We specifically study if the MWPC LETs and phase shifts can describe their empirical counterparts simultaneously---since both are essential for a theoretically sound PC with an ambition to quantitatively describe the nuclear force. Furthermore, we study the leading isospin-breaking effect in the one-pion exchange potential which is due to the pion mass splitting.

The Long and Yang PC differs from WPC in the $^1S_0$ partial wave by having promoted contact interactions and by treating sub-leading ($\nu>0$) interactions perturbatively. Ref.~\cite{Baru:2015ira} emphasizes that two-pion exchange contributions are expected to be important for describing LETs in the $^1S_0$ partial wave and it is unknown if this can be achieved with a perturbative inclusion. In the $\tS$ channel the only difference compared to WPC is that sub-leading interactions are treated perturbatively, meaning that our study in this channel specifically targets the feasibility of including two-pion exchange contributions perturbatively. Importantly, the EFT truncation error stemming from the truncated chiral expansion \cite{Ekstrom:2013kea,Schindler:2008fh,Furnstahl:2014xsa} is not taken into account in this study. The uncertainty in the predictions is instead estimated using the residual cutoff dependence~\cite{Griesshammer:2015osb}. It should be noted that Ref.~\cite{Gasparyan:2022isg} studied the cutoff dependence of the Long and Yang PC in the $^3P_0$ partial wave and found the appearance of so-called exceptional points in the cutoff domain for which the amplitude diverges. This is not a major concern in this study since the problematic regions in the vicinity of the exceptional points are extremely narrow.

The article is organized as follows. \Cref{sec:formalism} contains a brief overview of how \np{} scattering amplitudes are computed in the Long and Yang PC. In \cref{sec:LETs}, LETs are computed in the $^1S_0$ and $^3S_1$ partial waves at orders LO to \NNNLO{} and compared to similar studies \cite{Cohen:1998jr,Cohen:1999iaa,Ando:2011aa,Epelbaum:2003xx,Epelbaum:2012ua,Epelbaum:2015sha}. Finally, in \cref{sec:conclusions} we conclude and discuss prospects for future analyses carefully accounting for the EFT error and its effect on the predicted LETs. 

\section{Computing scattering amplitudes and phase shifts}\label{sec:formalism}
This section contains a brief summary of how \np{} scattering amplitudes are computed in \chiEFT{} using the Long and Yang PC developed in Refs.~\cite{Long:2012ve,PhysRevC.84.057001,PhysRevC.85.034002}. We consider a scattering process of an incoming neutron with kinetic energy $\Tl$ impinging on a proton. Specifically, we only investigate the two $S$-wave channels: $^1S_0$ and $\tS$. In \chiEFT{}, the \np{} potential gets contributions from both contact interactions of zero range and finite range interactions generated by pion exchanges. The potential contributions in the  Long and Yang PC are organized in chiral orders as \cite{Long:2012ve,PhysRevC.84.057001,PhysRevC.85.034002} 
\begin{equation}    
\begin{alignedat}{2}
    &\mathrm{LO:} &&V^{(0)} = V^{(0)}_{1\pi} + V^{(0)}_{\mathrm{ct}} \\
    &\mathrm{NLO:} &&V^{(1)} =  V^{(1)}_{\mathrm{ct}}\\
    &\mathrm{N}^2\mathrm{LO:}  &&V^{(2)} = V^{(2)}_{2\pi} + V^{(2)}_{\mathrm{ct}}\\
    &\mathrm{N}^3\mathrm{LO:} \ &&V^{(3)} = V^{(3)}_{2\pi} + V^{(3)}_{\mathrm{ct}}
\end{alignedat}
\label{eq:V_nonpert}
\end{equation}
where $V^{(\nu)}_{1\pi}$, $V^{(\nu)}_{2\pi}$ and $V^{(\nu)}_\mathrm{ct}$ denote one-pion exchange, two-pion exchange, and contact potentials respectively. The contact potentials are parameterized by LECs which need to be fixed using experimental data. For further details about the PC, we refer to Refs.~\cite{Long:2012ve,PhysRevC.84.057001,PhysRevC.85.034002,Thim:2024yks}.

The one-pion exchange potential enters at LO in the considered channels and reads \cite{Machleidt:2011zz}
\begin{equation}
    V^{(0)}_{1\pi} = -\frac{g^2_A}{4f^2_\pi} \frac{\left(\bm{\sigma}_1\cdot \bm{q}\right)\left(\bm{\sigma}_2\cdot \bm{q}\right)}{\bm{q}^2 + m^2_\pi} \Big[2I(I+1)-3\Big]
    \label{eq:OPE}
\end{equation}
where $\bm{\sigma}_i$, for $i=1,2$, denotes the spin operator for the respective nucleon, $I$ the total isospin, $\bm{p}$ ($\bm{p}'$) the ingoing (outgoing) relative $np$-momentum and $\bm{q} = \bm{p}'- \bm{p}$ the momentum transfer. The numerical values employed for the constants are: pion-nucleon axial coupling $g_A=1.29$ (including the Goldberger-Treiman discrepancy \cite{Goldberger:1958tr}), average pion mass $m_\pi=138.039$~MeV, and pion decay constant $f_\pi=92.1$~MeV. The leading charge-independence breaking (CIB) effect in the one-pion exchange potential comes from the pion mass splitting. This effect can be incorporated non-perturbatively at LO by modifying the \np{} one-pion exchange potential to
\begin{equation}
    V^{(0)}_{1\pi} = -\frac{g^2_A}{4f^2_\pi} \left(\bm{\sigma}_1\cdot \bm{q}\right)\left(\bm{\sigma}_2\cdot  \bm{q}\right)\left[-\frac{1}{\bm{q}^2 + m^2_{\pi^0}} + (-1)^{I+1}\frac{2}{\bm{q}^2 + m^2_{\pi^\pm}} \right],
    \label{eq:OPE_iso}
\end{equation}
where $m_{\pi^0} = 134.977$~MeV and $m_{\pi^\pm} = 139.570$~MeV denote the pion masses \cite{PDG_2022}.

The LO contact potential has contributions in the $^1S_0$ and $\tS$ channels which are given by
\begin{equation}
    V^{(0)}_\mathrm{ct} = C^{(0)}_{^1S_0} \hat{P}_{^1S_0} + C^{(0)}_{^3S_1} \hat{P}_{^3S_1},
\end{equation}
where $\hat{P}_X$ denotes the projector onto the given partial wave and the constants $C^{(0)}_X$ denote the corresponding LO LECs. The superscript indicates that these are the first contributions to the LECs: $C_{^1S_0}$ and $C_{^3S_1}$ which also receive contributions at sub-leading orders since the higher order potentials are added perturbatively \cite{Long:2012ve,Contessi:2017rww}. For the two-pion exchange, we use expressions computed using dimensional regularization \cite{Machleidt:2011zz} and for the sub-leading two-pion exchange ($V^{(3)}_{2\pi}$) we employ $c_i$ LECs: $c_1 = -0.74$~GeV$^{-1}$, $c_3 = -3.61$~GeV$^{-1}$ and $c_4 = 2.44$~GeV$^{-1}$ from the Roy-Steiner analysis of pion-nucleon scattering amplitudes at order $Q^2$ presented in Ref.~\cite{Siemens:2016jwj}. Explicit expressions for all potentials are documented in the Appendix of Ref.~\cite{Thim:2024yks}.

The \np{} scattering amplitude at order $\nu$ is denoted $T^{(\nu)}$, analogous to the notation for the potentials. The \LO{} amplitude is computed non-perturbatively by solving the Lippmann–Schwinger equation with the LO potential
\begin{equation}
    \TLO = \VLO + \VLO G^+_0\TLO,
    \label{eq:LS_op}
\end{equation}
where $G^+_0 = \left(E - H_0 + i\epsilon\right)^{-1}$ is the free resolvent, $H_0 = \bm{p}^2/m_N$ the free Hamiltonian, $m_N$ the nucleon mass and $E=k^2/m_N$ the center-of-mass energy corresponding to relative momentum $k$. Sub-leading corrections to the scattering amplitude are computed using distorted wave perturbation theory \cite{Long:2007vp,Peng:2020nyz} using the equations \cite{Thim:2024yks}
\begin{align}
    \TNLO &= \Omega^\dagger_-\VNLO\Omega_+ \label{eq:TNLO}\\
    \TNNLO &= \Omega^\dagger_-\left(\VNNLO  +   \VNLO G^+_1 \VNLO\right) \Omega_+\label{eq:TN2LO}\\
    \TNNNLO &= \Omega^\dagger_-\Big(\VNNNLO +   \VNNLO G^+_1\VNLO +  \VNLO G^+_1\VNNLO + \nonumber\\ &+ \VNLO G^+_1\VNLO G^+_1\VNLO \Big)\Omega_+, \label{eq:TN3LO}
\end{align}
where we have introduced the operators $\Omega_+ = \mathds{1} + G^+_0\TLO$, $\Omega^\dagger_- = \mathds{1} + \TLO G^+_0$ and $G^+_1 = \Omega_+ G^+_0$. The partial wave scattering amplitudes are computed numerically using matrix discretization \cite{Haftel:1970zz} by expressing \cref{eq:LS_op,eq:TNLO,eq:TN2LO,eq:TN3LO} in a partial wave basis $\ket{p,\l s j}$. Here, $p=\abs{\bm{p}}$, and $\l$, $s$, $j$ are the quantum numbers associated with the \np{} orbital angular momentum, spin, and total angular momentum, respectively. The partial-wave Lippmann–Schwinger equation reads 
\begin{align}
    T^{(0)sj}_{\l'\l}(p', p) = V^{(0)sj}_{\l'\l}(p',p) + \sum_{\l''}\int_0^\infty dq \ q^2  V^{(0)sj}_{\l'\l''}(p',q) \frac{m_N}{k^2-q^2 + i\epsilon} T^{(0)sj}_{\l''\l}(q,p),
    \label{eq:LS_pw}
\end{align}
where $k=\sqrt{m_N E}$ denote the on-shell relative momentum corresponding to $\Tl$ \cite{Glockle}. The notation $T^{(\nu) sj}_{\l'\l}(p',p) \equiv \bra{p',\l' s j}T^{(\nu)}\ket{p,\l s j}$ is used for amplitudes and potentials.
The partial wave projected potentials are regulated using the transformation
\begin{equation}
    V^{(\nu)sj}_{\l'\l}(p',p) \to \exp\left(-\frac{p'^6}{\Lambda^6}\right)V^{(\nu)sj}_{\l'\l}(p',p)\exp\left(-\frac{p^6}{\Lambda^6}\right)
\end{equation}
where $\Lambda$ is referred to as the momentum cutoff. 

The contributions to the scattering amplitude at each order, $T^{(\nu) sj}_{\l'\l}$, can now be computed for $\nu \leq 3$ in each scattering channel. The total on-shell amplitude ($p'=p=k$) is obtained by summing the contributions to the desired order according to
\begin{equation}
    T^{sj}_{\l'\l}(\pon,\pon) = T^{(0)sj}_{\l'\l}(\pon,\pon) + T^{(1)sj}_{\l'\l}(\pon,\pon) + T^{(2)sj}_{\l'\l}(\pon,\pon) + \dots \label{eq:T-sum}
\end{equation}
Using the relation between the scattering amplitude and $S$-matrix,
\begin{equation}
    S^{sj}_{\l'\l} = \delta_{\l'\l}- i\pi \pon \mN T^{sj}_{\l'\l},
    \label{eq:ST_pw}
\end{equation}
phase shift contributions can be computed at each chiral order by expanding \cref{eq:ST_pw} and matching chiral orders. This is described in Appendix \ref{app:pert_phase} and yields expansions for the corresponding shifts in chiral orders analogous to \cref{eq:T-sum}
\begin{equation}
    \delta(\pon) = \delta^{(0)}(\pon) + \delta^{(1)}(\pon) + \delta^{(2)}(\pon) +\dots \label{eq:phase-sum}
\end{equation}
In the next section, the scattering amplitudes are utilized to compute both phase shifts and LETs.

\section{Low-energy theorems for effective range parameters}\label{sec:LETs}
The $S$-wave effective range function, $F(k)$, can be expressed in terms of the $S$-wave phase shift, $\delta(k)$, and possesses the property of being analytic in $k^2$ near the origin giving the ERE
\begin{equation}
    F(\pon)  \equiv \pon \cot{\delta}\left(\pon\right) = -\frac{1}{a} + \frac{1}{2} r \pon^2 + v_2 \pon^4 + v_3 \pon^6 + v_4 \pon^8 + \mathcal{O}\left(\pon^{10}\right).
    \label{eq:F_ERE}
\end{equation}
The ERE parameters $a$ and $r$ are called the scattering length and effective range, while $v_2,v_3$ and $v_4$ are referred to as shape parameters. The presence of pions in \chiEFT{} restricts the radius of convergence for the ERE to $\pon < m_\pi/2 \approx 69$~MeV corresponding to the first left-hand cut in the scattering amplitude caused by one-pion exchange~\cite{Baru:2015ira}. Hence, the ERE converges for $\Tl \lesssim 10$~MeV, in what is referred to as the low-energy regime.

To compute LETs, the unknown LECs must first be inferred from either phase shifts \cite{Stoks:1993tb}, the first few empirical effective range parameters, or a mix of both. The effective range function in \cref{eq:F_ERE} can then be computed, and the remaining ERE parameters not used to calibrate the LECs are predictions governed by the long-range part of the interaction (LETs) \cite{Cohen:1998jr}. In the coming sections, we will compute LETs in the $^1S_0$ and $^3S_1$ partial waves where the scattering amplitudes and the effective range function are computed perturbatively beyond LO.

\subsection{The $^1S_0$ partial wave}
We first consider the $^1S_0$ partial wave and thus suppress its associated quantum numbers from the notation. For uncoupled \np{} scattering channels, it is convenient to express the effective range function directly in terms of the on-shell scattering amplitude, $T(\pon,\pon)$,
using the relation \cite{Taylor72}
\begin{equation}
    T(\pon,\pon) = -\frac{2}{\pi \mN}\frac{1}{F(\pon) - i \pon}.
    \label{eq:T-F}
\end{equation}
By treating sub-leading amplitudes as perturbations to $T^{(0)}$, Taylor expanding \cref{eq:T-F} and keeping terms to order $\nu=3$ one obtains
\begin{align}
    F(\pon) - i\pon &= -\frac{2}{\pi \mN T^{(0)}} \Bigg[ 1-\frac{T^{(1)}}{T^{(0)}}  + \left(\left[\frac{T^{(1)}}{T^{(0)}}\right]^2- \frac{T^{(2)}}{T^{(0)}}\right) + \nonumber \\ &+\left( 2\frac{T^{(1)} T^{(2)}}{\big(T^{(0)}\big)^2} - \frac{T^{(3)}}{T^{(0)}} - \left[\frac{T^{(1)}}{T^{(0)}}\right]^3\right) + \mathcal{O}\left(\frac{Q^4}{\Lambda^4_b}\right)\Bigg]. \label{eq:F_expansion}
\end{align}
The effective range function can be written in terms of contributions at each chiral order, analogous to the scattering amplitude
\begin{equation}
    F(\pon) = F^{(0)}(\pon)+F^{(1)}(\pon)+F^{(2)}(\pon)+ \dots
    \label{eq:F_chiral}
\end{equation}
These contributions are identified in \cref{eq:F_expansion} and read
\begin{align}
    F^{(0)}(\pon)  &= -\frac{2}{\pi \mN T^{(0)}} + i\pon \label{eq:F_T_0}\\
    F^{(1)}(\pon) &= \frac{2}{\pi \mN T^{(0)}} \frac{T^{(1)}}{T^{(0)}} \label{eq:F_T_1}\\
    F^{(2)}(\pon) &= -\frac{2 }{\pi \mN T^{(0)}}\left(\left[\frac{T^{(1)}}{T^{(0)}}\right]^2- \frac{T^{(2)}}{T^{(0)}}\right) \label{eq:F_T_2}\\
    F^{(3)}(\pon) &= -\frac{2}{\pi \mN T^{(0)}}\left( 2\frac{T^{(1)} T^{(2)}}{\big(T^{(0)}\big)^2} - \frac{T^{(3)}}{T^{(0)}} - \left[\frac{T^{(1)}}{T^{(0)}}\right]^3\right). \label{eq:F_T_3}
\end{align}
Note that in a theory without pions, the renormalized LO amplitude possesses the property that $\mathrm{Re}\big\{\left(T^{(0)}\right)^{-1}\big\}$ is a momentum independent constant and all coefficients but $a$ in the ERE will be zero  by \cref{eq:F_T_0}. This illustrates the fact that non-zero ERE parameters beyond $a$ at LO can be attributed to the presence of a long-range force, in our case the one-pion exchange.

\begin{figure}
    \centering
    \includegraphics[width=\textwidth]{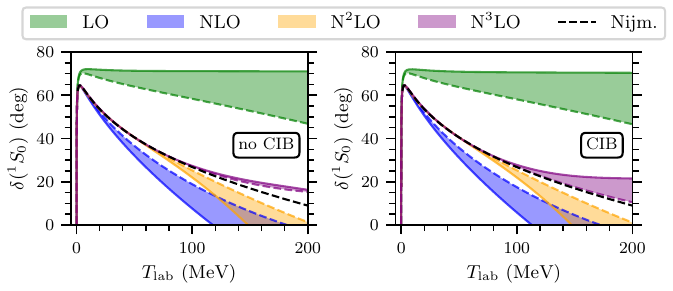}
    \caption{Phase shifts in the $^1S_0$ partial wave as a function of laboratory scattering energy for the LEC calibration with corresponding LETs in \cref{tab:ERE_1S0_phase_fit}. The left (right) panel shows results without (with) CIB, corresponding to the expression in \cref{eq:OPE} (\cref{eq:OPE_iso}) for the one-pion exchange potential. The bands indicate the envelope of the variation due to the two different cutoff values; 500 MeV (dashed line) and 2500 MeV (solid line). The black dashed line shows phase shifts from the Nijmegen analysis \cite{Stoks:1993tb}.}
    \label{fig:1S0_fit_phase}
\end{figure}

\begin{table}
    \caption{LETs for the $^1S_0$ partial wave, both without and with the leading CIB, corresponding to the phase shifts in \cref{fig:1S0_fit_phase}. The ERE parameters marked with a star ($*$) are used in the inference and are not predictions. The quoted errors only stem from the numerical extraction of the ERE parameters from the perturbatively computed effective range function. Empirical ERE parameters are also shown.}
\centering
\begingroup
\begin{tabular}{l|ccccc}
    \toprule
    $^1S_0$ partial wave & $a$ [fm] & $r$ [fm] & $v_2$ [fm$^3$]& $v_3$ [fm$^5$] & $v_4$ [fm$^7$]\\
    \hline
    Empirical (Ref.~\cite{NavarroPerez:2014ovp}) & $-23.735(16)$ & 2.68(3) & $-0.48(2)$ & 3.9(1) & $-19.6(5)$ \\
    \hline
    \underline{$\Lambda = 500$~MeV, (no CIB)} \\
    LO
     & $*$ & $1.71(0)$ & $-1.77(0)$ & $8.54(0)$ & $-47.0(3)$ \\
    NLO
     & $*$ & $*$ & $-0.64(0)$ & $4.79(0)$ & $-29.9(2)$ \\
    \NNLO
     & $*$ & $2.72(0)$ & $-0.71(0)$ & $5.05(0)$ & $-29.3(2)$ \\
    \NNNLO
     & $*$ & $2.69(0)$ & $-0.66(0)$ & $5.42(0)$ & $-31.0(2)$ \\
    \hline
    
    \underline{$\Lambda = 2500$~MeV, (no CIB)} \\

    \LO
    & $*$ & $1.49(0)$ & $-2.06(0)$ & $9.34(0)$ & $-50.7(3)$ \\
    \NLO
    & $*$ & $*$ & $-0.55(0)$ & $4.70(0)$ & $-30.1(2)$ \\
    \NNLO
    & $*$ & $2.75(0)$ & $-0.75(0)$ & $4.80(0)$ & $-28.1(2)$ \\
    \NNNLO
    & $*$ & $2.70(0)$ & $-0.69(0)$ & $5.52(0)$ & $-30.6(5)$ \\
    \botrule
    \underline{$\Lambda = 500$~MeV, (CIB)} \\ 
    LO                                                                             
   & $*$ & $1.68(0)$ & $-1.55(0)$ & $6.63(0)$ & $-31.64(8)$ \\            
  NLO                                                                            
   & $*$ & $*$ & $-0.45(0)$ & $3.42(0)$ & $-18.95(8)$ \\            
  \NNLO                                                                           
   & $*$ & $2.70(0)$ & $-0.55(0)$ & $3.77(0)$ & $-18.8(2)$ \\             
  \NNNLO                                                                           
   & $*$ & $2.68(0)$ & $-0.50(0)$ & $4.02(0)$ & $-19.8(2)$ \\

\hline
    \underline{$\Lambda = 2500$~MeV, (CIB)} \\
LO                                                                             
   & $*$ & $1.47(0)$ & $-1.81(0)$ & $7.27(0)$ & $-34.23(8)$ \\            
  NLO                                                                            
   & $*$ & $*$ & $-0.36(0)$ & $3.35(0)$ & $-19.13(8)$ \\            
  \NNLO                                                                           
   & $*$ & $2.72(0)$ & $-0.59(0)$ & $3.56(0)$ & $-17.7(3)$ \\             
  \NNNLO                                                                           
   & $*$ & $2.67(0)$ & $-0.52(0)$ & $4.26(2)$ & $-20.0(7)$ \\ 
   \botrule
\end{tabular}
\endgroup
\label{tab:ERE_1S0_phase_fit}
\end{table}

To compute phase shifts and LETs, the unknown LECs first need to be inferred. In this study, we neglect both the truncation error associated with the \chiEFT{} expansion and the uncertainties associated with the calibration data during the inference, similar to earlier studies \cite{Cohen:1998jr,Cohen:1999iaa,Epelbaum:2012ua,Epelbaum:2015sha}. Instead, we utilize that adding higher chiral orders improves the high-energy description. Consequently, we successively include higher-energy data in the calibration as the chiral order is increased. A measure of the theoretical uncertainty is provided by doing the calculations using different momentum cutoffs $\Lambda=500$~MeV and $\Lambda=2500$~MeV, where the residual cutoff dependence is expected to indicate the effect of excluded terms in the chiral expansion \cite{Griesshammer:2015osb}. 

To gauge the leading effect of CIB in the \np{} interaction we will consider two versions of the one-pion exchange potential at LO: the isospin symmetric (\cref{eq:OPE}) and the isospin breaking (\cref{eq:OPE_iso}).

We wish to obtain LETs that describe empirical ERE parameters while the predicted phase shifts show a good description of their empirical counterparts. To achieve this, we employ both empirical phase shifts and ERE parameters as data to calibrate the LECs, where the latter are shown in the first row of \cref{tab:ERE_1S0_phase_fit} \cite{NavarroPerez:2014ovp}. Note that mainly phase shifts are used at the higher orders, to ensure a satisfactory description of the empirical phase shifts over the wide energy interval. The LEC at LO, $C^{(0)}_{^1S_0}$, is calibrated to reproduce the scattering length. The two LECs at NLO are calibrated to reproduce both $a$ and $r$. At \NNLO{} the three LECs are calibrated using $a$, as well as the Nijmegen phase shifts \cite{Stoks:1993tb} at $T_\mathrm{lab}=25$~MeV and 50~MeV. At \NNNLO{} the four LECs are inferred from $a$ as well as the Nijmegen phase shifts at $T_\mathrm{lab} = 5$~MeV, 50~MeV, and $75$~MeV. LETs are computed by performing a least squares fit of the ERE polynomial (\cref{eq:F_ERE}) to the effective range function computed to the desired order: 
$F^{(0)}(\pon)$ at LO, $F^{(0)}(\pon) + F^{(1)}(\pon)$ at NLO, and so on. To estimate the error in the LETs, a series of least squares fits are performed in momentum intervals $(p_l,p_h)$, where $0.6\ \text{MeV} <p_l < 1.7 \ \text{MeV}$ and $30\ \text{MeV} <p_h<68 \ \text{MeV} $ for varying polynomial orders.

The predicted phase shifts without (with) CIB are presented in the left (right) panel of \cref{fig:1S0_fit_phase}. The phase shifts show a clear order-by-order convergence, and at \NNNLO{} they agree excellently with the Nijmegen phase shift up to $\Tl \approx 100$~MeV. We also observe a minimal difference in the phase shifts with and without CIB and a decreasing cutoff dependence as the chiral order is increased. The LETs corresponding to \cref{fig:1S0_fit_phase} are presented in \cref{tab:ERE_1S0_phase_fit}. Contrary to the almost non-visible difference in the phase shifts, a substantial improvement is observed for the LETs when accounting for CIB. The presented uncertainties in the LETs only stem from the numerical extraction of the ERE parameters from the perturbatively computed effective range function. These errors are negligible for all ERE parameters except $v_4$.

Without considering CIB, the obtained LETs for LO and NLO are consistent with similar studies~\cite{Epelbaum:2012ua,Epelbaum:2015sha}, whose results are summarized in \cref{tab:ERE_1S0_ref} in Appendix \ref{app:tables}.
The LETs improve considerably from LO to NLO, i.e. when including the contact interaction $D^{(0)}_{^1S_0}\left(p^2+p'^2\right)$ \cite{Long:2012ve,Thim:2024yks}, while no further improvements are observed at \NNLO{} and \NNNLO. Thus, convergence to the empirical values of the ERE parameters is not observed as the chiral order is increased. Still, we note that the LETs are significantly improved compared to KSW counting \cite{Kaplan:1998tg,Kaplan:1998we,Cohen:1998jr}, see \cref{tab:ERE_1S0_ref}. 

When considering CIB, the LETs at all orders improve drastically. The largest improvement is again observed when going from LO to NLO. However, there are now further improvements beyond NLO, especially for $v_2$ and $v_3$. At \NNNLO{}, the LETs for $\Lambda=500$~MeV are consistent with the empirical ERE parameters within the quoted errors. For $\Lambda=2500$~MeV there is still a small tension for $v_2$ and $v_3$, but this is without considering any EFT error. We note that the LETs obtained at \NNNLO{} are comparable to the ones obtained in WPC \cite{Epelbaum:2003xx}, see \cref{tab:ERE_1S0_ref}.

We now want to explore the effect of employing an alternate scheme to infer the LECs, gauging how the choice of calibration data impacts the results. Instead of phase shifts, we will mainly use the empirical ERE parameters to calibrate the LECs. We calibrate the LECs to reproduce $\{a\}$, $\{a,r\}$ and $\{a,r,v_2\}$ for LO to \NNLO{} respectively. At \NNNLO{} there is one additional LEC compared to \NNLO, and $v_3$ can be included in the calibration data. However, we find no set of LECs that can reproduce $\{a,r,v_2,v_3\}$ simultaneously. We find, however, that it is possible to achieve a fit by slightly adjusting the values of some empirical ERE parameters, which hints at the fact that accounting for uncertainties can resolve this apparent issue. At \NNNLO{} in this calibration scheme, the LECs are instead tuned to reproduce $a$ and the phase shifts at $T_\mathrm{lab} = 5,50$ and $75$~MeV.

\begin{figure}
    \centering
    \includegraphics[width=\textwidth]{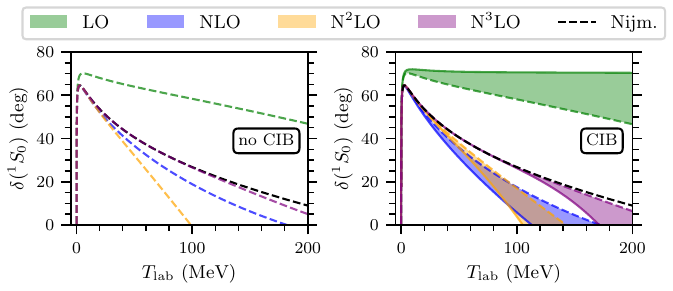}
    \caption{Phase shifts in the $^1S_0$ partial wave as a function of laboratory scattering energy for the LEC calibration with corresponding LETs in \cref{tab:ERE_1S0_phase_fit_vary}. The left (right) panel shows results without (with) CIB, corresponding to the expression in \cref{eq:OPE} (\cref{eq:OPE_iso}) for the one-pion exchange potential. The left panel shows the predicted phase shifts for $\Lambda=500$~MeV. The bands in the right panel indicate the envelope of the variation due to the two different cutoff values; 500 MeV (dashed line) and 2500 MeV (solid line). The black dashed line shows phase shifts from the Nijmegen analysis \cite{Stoks:1993tb}.}
    \label{fig:phase_1S0_vary}
\end{figure}

When considering CIB, this calibration works well for $\Lambda=500$~MeV and $\Lambda=2500$~MeV and the resulting phase shifts are shown in the right panel of \cref{fig:phase_1S0_vary}. Without considering CIB, reproducing $\{a,r,v_2\}$ exactly at \NNLO{} is not possible for $\Lambda=2500$~MeV. One potential explanation can be that the LO phase shift for $\Lambda=2500$~MeV exhibits a greater overestimation of the Nijmegen phase shift compared to the $\Lambda=500$~MeV phase shift (see \cref{fig:1S0_fit_phase}), thus necessitating a larger correction from the sub-leading orders. Consequently, only phase shifts for $\Lambda=500$~MeV are shown in the left panel of \cref{fig:phase_1S0_vary}. We see that the phase shifts in \cref{fig:phase_1S0_vary} only exhibit minor differences compared to the phase shifts in \cref{fig:1S0_fit_phase}, with the distinction that \NNLO{} perform notably worse.

The LETs corresponding to \cref{fig:phase_1S0_vary} are presented in \cref{tab:ERE_1S0_phase_fit_vary}. When considering CIB, the LETs are similar to those obtained in \cref{tab:ERE_1S0_phase_fit}. We again note that the LETs show no tension with the empirical ERE parameters for $\Lambda=500$~MeV, while a slight tension for $v_3$ persists at \NNNLO{} for $\Lambda=2500$~MeV. When not considering CIB, the LETs at \NNLO{} perform worse than the ones at \NLO{} as well as the \NNLO{} LETs from \cref{tab:ERE_1S0_phase_fit}. It is clear that including $v_2$ in the calibration data is problematic when not considering CIB. This is likely a contributing factor to the failure of calibrating the LECs for $\Lambda=2500$~MeV.

\begin{table}
    \caption{LETs for the $^1S_0$ partial wave, both without and with the leading CIB, corresponding to the phase shifts in \cref{fig:phase_1S0_vary}. The parameters marked with a star ($*$) are used in the inference and are not predictions. The quoted errors only stem from the numerical extraction of the ERE parameters from the perturbatively computed effective range function. Empirical ERE parameters are also shown.}
\centering
\begingroup
\begin{tabular}{l|ccccc}
    \toprule
    $^1S_0$ partial wave & $a$ [fm] & $r$ [fm] & $v_2$ [fm$^3$]& $v_3$ [fm$^5$] & $v_4$ [fm$^7$]\\
    \hline
    Empirical (Ref.~\cite{NavarroPerez:2014ovp}) & $-23.735(16)$ & 2.68(3) & $-0.48(2)$ & 3.9(1) & $-19.6(5)$ \\
    \hline
    \underline{$\Lambda=500$~MeV (no CIB)} \\
    LO
    & $*$ & $1.71(0)$ & $-1.77(0)$ & $8.54(0)$ & $-47.0(3)$ \\
    NLO
    & $*$ & $*$ & $-0.64(0)$ & $4.79(0)$ & $-29.9(2)$ \\
    \NNLO
    & $*$ & $*$ & $*$ & $5.89(0)$ & $-32.3(2)$ \\
    \NNNLO
    & $*$ & $2.69(0)$ & $-0.61(0)$ & $5.51(0)$ & $-30.9(2)$ \\
    \hline
    \underline{$\Lambda=500$~MeV (CIB)} \\
     LO                                                     & $*$ & $1.68(0)$ & $-1.55(0)$ & $6.63(0)$ & $-31.64(8)$ \\            
 NLO                                                                            
  & $*$ & $*$ & $-0.45(0)$ & $3.42(0)$ & $-18.95(8)$ \\            
  \NNLO                                                                           
 & $*$ & $*$ & $*$ & $4.05(0)$ & $-19.7(2)$ \\             
 \NNNLO                                                                           
  & $*$ & $2.68(0)$ & $-0.49(0)$ & $4.04(0)$ & $-19.7(2)$ \\       
    \hline
    \underline{$\Lambda=2500$~MeV (CIB)} \\
     LO                                                                            
   & $*$ & $1.47(0)$ & $-1.81(0)$ & $7.27(0)$ & $-34.23(8)$ \\            
  \NLO                                                                            
   & $*$ & $*$ & $-0.36(0)$ & $3.35(0)$ & $-19.13(8)$ \\            
  \NNLO                                                                           
   & $*$ & $*$ & $*$ & $4.08(0)$ & $-19.3(3)$ \\             
  \NNNLO                                                                          
   & $*$ & $2.68(0)$ & $-0.49(0)$ & $4.17(0)$ & $-19.5(8)$ \\
    \botrule
\end{tabular}
\endgroup
\label{tab:ERE_1S0_phase_fit_vary}
\end{table}

The analysis in the $^1S_0$ partial wave can be concluded as follows. In all cases considered, the cutoff $\Lambda=500$~MeV shows a better convergence for both phase shifts and LETs compared to $\Lambda=2500$~MeV. This can likely be explained by the lower cutoff producing a larger effective range at LO, requiring smaller corrections from the sub-leading orders. When not considering CIB in the one-pion exchange potential, a clear discrepancy between LETs and empirical ERE parameters is observed while phase shifts show a good convergence to their empirical counterparts up to  $\Tl\approx100$~MeV (\cref{fig:1S0_fit_phase} and \cref{tab:ERE_1S0_phase_fit}). When considering CIB, a substantial improvement for the LETs is observed while the phase shifts again show a good convergence up to $\Tl\approx100$~MeV. We can conclude that CIB is an important effect to include in the $^1S_0$ partial wave to be able to describe both phase shifts and LETs. The alternate calibration of the LECs (see \cref{fig:phase_1S0_vary} and \cref{tab:ERE_1S0_phase_fit_vary}) strengthens this conclusion since the results when considering CIB only show small differences in both LETs and phase shifts compared to the first calibration.

\subsection{The $^3S_1$ partial wave}
We move on to the $^3S_1$ partial wave, which is part of the coupled scattering channel denoted $\tS$. The mixing of states with different $\l$ means that the relation between the scattering amplitude and the effective range function in \cref{eq:T-F} cannot be used directly. A general method to numerically extract ERE parameters in coupled channels utilizing the scattering amplitudes was developed in Ref.~\cite{PavonValderrama:2005ku}. However, for perturbative calculations, this method is not directly applicable. Instead, we express the effective range function in terms of the Blatt and Biedenharn (BB) \cite{Blatt:1952zz} eigen phase shift in the $^3S_1$ partial wave, denoted $\dtS(k)$ \cite{PavonValderrama:2005ku}. The phase shift contributions at each chiral order, $\dtS^{(\nu)}(\pon)$, are computed from the scattering amplitudes as described in Appendix \ref{app:pert_phase}.
The contributions to the effective range function at each chiral order are computed by expanding \cref{eq:F_ERE}, yielding
\begin{align}
    F^{(0)}(\pon) &= \pon \cot\big(\dtS^{(0)}\big), \nonumber \\
    F^{(1)}(\pon) &= \pon \frac{d\cot\big(\dtS^{(0)}\big)}{d\delta}\times \dtS^{(1)}, \nonumber \\
    F^{(2)}(\pon) &= \pon \left[\frac{d\cot\big(\dtS^{(0)}\big)}{d\delta}\times \dtS^{(2)} + \frac{1}{2}\frac{d^2\cot\big(\dtS^{(0)}\big)}{d\delta^2}\times \big(\dtS^{(1)}\big)^2\right], \label{eq:F_phase}\\
    F^{(3)}(\pon) &= \pon \Bigg[\frac{d\cot\big(\dtS^{(0)}\big)}{d\delta}\times \dtS^{(3)} + \frac{d^2\cot\big(\dtS^{(0)}\big)}{d\delta^2}\times \dtS^{(1)}\dtS^{(2)} +\nonumber \\ 
    &+\frac{1}{6}\frac{d^3\cot\big(\dtS^{(0)}\big)}{d\delta^3}\times \big(\dtS^{(1)}\big)^3 \Bigg].\nonumber
\end{align}
The Long and Yang PC dictates that the potential contribution at NLO is zero in the $\tS$ channel \cite{PhysRevC.85.034002} which leads to simplifications in the above equations since $\dtS^{(1)} = 0$. Computing the effective range function in terms of the phase shifts can also be done in the $^1S_0$ partial wave and we confirmed that it gives the same result as using \cref{eq:F_T_0,eq:F_T_1,eq:F_T_2,eq:F_T_3}.

\begin{figure}
    \centering
    \includegraphics[width=\textwidth]{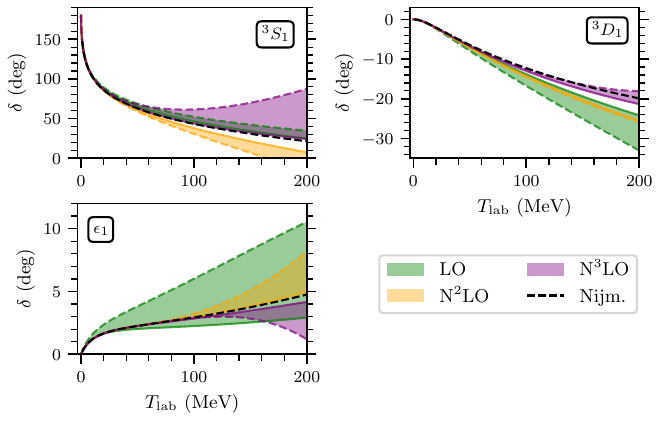}
    \caption{Phase shifts in the $\tS$ channel in the BB convention \cite{Blatt:1952zz} as a function of scattering energy for the LEC calibration with corresponding LETs in \cref{tab:ERE_3S1_phase_fit}. CIB is not considered, and the one-pion exchange expression in \cref{eq:OPE} is employed. The bands indicate the envelope of the cutoff variation; $\Lambda=500$~MeV (dashed line) and $\Lambda=2500$~MeV (solid line). The black dashed line shows phase shifts from the Nijmegen analysis \cite{Stoks:1993tb} in the  BB convention.}
    \label{fig:phase_3S1}
\end{figure}
\begin{table}
    \caption{LETs in the $^3S_1$ partial wave without considering CIB. The parameters marked with a star ($*$) are used in the inference and are not predictions. The quoted errors only stem from the numerical extraction of the ERE parameters from the effective range function. Empirical ERE parameters are also shown.}
\centering
\begingroup
\begin{tabular}{l|ccccc}
    \toprule
    $^3S_1$ partial wave & $a$ [fm] & $r$ [fm] & $v_2$ [fm$^3$]& $v_3$ [fm$^5$] & $v_4$ [fm$^7$]\\
    \hline
     Empirical (Ref.~\cite{NavarroPerez:2014ovp}) & 5.42 &   1.75 &0.045& 0.67 &$-3.94$ \\
    \hline
    \underline{$\Lambda = 500$~MeV} \\
   
  LO
 & $*$ & $1.58(0)$ & $-0.10(0)$ & $0.89(0)$ & $-5.5(2)$ \\
\NNLO
 & $*$ & $*$ & $0.14(0)$ & $0.80(0)$ & $-4.2(2)$ \\
\NNNLO
 & $*$ & $*$ & $-0.06(0)$ & $0.46(0)$ & $-3.7(2)$ \\
    \hline
    \underline{$\Lambda = 750$~MeV} \\
   
    LO
    & $*$ & $1.69(0)$ & $0.01(0)$ & $0.77(0)$ & $-4.5(4)$ \\
    \NNLO
    & $*$ & $*$ & $0.10(0)$ & $0.77(0)$ & $-4.2(4)$ \\
    \NNNLO
    & $*$ & $*$ & $0.01(0)$ & $0.62(0)$ & $-4.0(4)$ \\
    \hline
    
    \underline{$\Lambda = 1000$~MeV} \\
    LO
    & $*$ & $1.69(0)$ & $0.01(0)$ & $0.77(0)$ & $-4.6(4)$ \\
    \NNLO
    & $*$ & $*$ & $0.09(0)$ & $0.75(0)$ & $-4.2(7)$ \\
    \NNNLO
    & $*$ & $*$ & $0.04(0)$ & $0.67(0)$ & $-4.0(4)$ \\
    \hline
    \underline{$\Lambda = 2500$~MeV} \\
    LO
    & $*$ & $1.66(0)$ & $-0.01(0)$ & $0.79(0)$ & $-4.7(2)$ \\
    \NNLO
    & $*$ & $*$ & $0.09(0)$ & $0.74(0)$ & $-4.2(7)$ \\
    \NNNLO
    & $*$ & $*$ & $0.04(0)$ & $0.67(2)$ & $-4.0(9)$ \\
 
    \botrule
\end{tabular}
\endgroup
\label{tab:ERE_3S1_phase_fit}
\end{table}

\begin{figure}
    \centering
    \includegraphics[width=\textwidth]{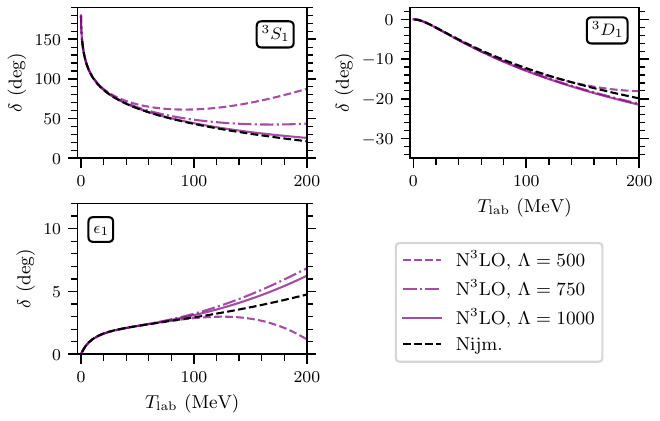}
    \caption{Predicted phase shifts in the $\tS$ channel in the BB convention \cite{Blatt:1952zz} as a function of scattering energy at \NNNLO{} for the LEC calibration with corresponding LETs in \cref{tab:ERE_3S1_phase_fit}. CIB is not considered, and the one-pion exchange expression in \cref{eq:OPE} is employed. The purple dashed, dot-dashed, and solid lines correspond to momentum cutoffs $\Lambda=500$~MeV, $\Lambda=750$~MeV, and $\Lambda=1000$~MeV respectively. The black dashed line shows phase shifts from the Nijmegen analysis \cite{Stoks:1993tb} in the  BB convention.}
    \label{fig:phase_3S1_N3LO_vary}
\end{figure}

The LECs in the $\tS$ channel are calibrated using data that ensures a good description of both phase shifts and ERE parameters---the same principle as used in the $^1S_0$ partial wave. The empirical values for the ERE parameters \cite{NavarroPerez:2014ovp} used in the inference are displayed in \cref{tab:ERE_3S1_phase_fit}. At LO the only LEC, $C^{(0)}_{^3S_1}$, is inferred by reproducing the $^3S_1$ scattering length. The NLO contribution is zero and contains no LECs, but at \NNLO{} there are three LECs: $C^{(1)}_{^3S_1}$, $D^{(0)}_{^3S_1}$ and $D^{(0)}_{SD}$. These are inferred by reproducing the scattering length ($a$), effective range ($r$), and the mixing angle $(\epsilon_{1})$ from the Nijmegen analysis at $T_\mathrm{lab}=50$~MeV \cite{Stoks:1993tb}. The three LECs at \NNNLO: $C^{(2)}_{^3S_1}$, $D^{(1)}_{^3S_1}$ and $D^{(1)}_{SD}$ are perturbative corrections to the LECs at \NNLO{} and are fixed using the same data as at \NNLO. The LECs are inferred using momentum cutoffs $\Lambda=500,750,1000$ and 2500~MeV, where the intermediate cutoffs are used to get a better understanding of the residual cutoff dependence.

The effect of CIB was also considered, but numerical computations showed that it is negligible in this partial wave. Thus, we only present results without CIB, i.e. employing the one-pion exchange expression in \cref{eq:OPE}. The resulting phase shifts for the extreme values of the cutoffs $\Lambda=500$~MeV and $\Lambda=2500$~MeV are shown in \cref{fig:phase_3S1}. The residual cutoff dependence should decrease as the chiral order increases. This is not observed at \NNNLO{}, since the phase shifts for $\Lambda=500$~MeV deviate significantly from both the Nijmegen and $\Lambda=2500$~MeV phase shifts. However, the results for $\Lambda=2500$~MeV show an excellent agreement with both empirical phase shifts, empirical ERE parameters, and similar studies \cite{Epelbaum:2003xx,Epelbaum:2012ua}.

LETs are computed from the effective range function using the same method as in the $^1S_0$ partial wave and the results are shown in \cref{tab:ERE_3S1_phase_fit}. The LETs show an expected convergence order-by-order, where $v_2$ is the exception, and especially for $\Lambda=500$~MeV. This can likely be attributed to the small numerical value of $v_2$, which makes it harder to extract reliably---a fact that is also reflected in a larger relative error between extractions from different high-precision potentials compared to $v_3$ and $v_4$ \cite{NavarroPerez:2014ovp}.

Both the LETs and phase shifts show a large residual cutoff dependence at \NNNLO{} reflecting the fact that the $\Lambda=500$~MeV result is deviating significantly from the results with higher cutoffs. The cutoff dependence is further illustrated in \cref{fig:phase_3S1_N3LO_vary} which shows predicted phase shifts at \NNNLO{} for the lowest cutoffs $\Lambda=500,750,$ and 1000~MeV. It is observed that the deviation of the predicted phase shifts compared to the Nijmegen analysis increases as the cutoff decreases, which is consistent with the observed cutoff variation of LETs in \cref{tab:ERE_3S1_phase_fit}. The residual cutoff dependence at \NNNLO{} seems to be under control if the cutoff is chosen sufficiently large, $\Lambda\gtrsim750$~MeV. This large cutoff dependence for low cutoffs can have several explanations. It can be an indication that the convergence radius for the perturbative computations in the $\tS$ channel is lower than expected, and the residual cutoff dependence signals that the perturbative series starts to diverge. It can also be an effect of a sub-optimal calibration procedure employed to fix the LECs. Another culprit can be the $c_i$ LECs parameterizing the strength of the sub-leading two-pion exchange. In Ref.~\cite{Long:2012ve} it was observed that some phase shifts at \NNNLO{} can be very sensitive to the employed values of the $c_i$ LECs, in line with the findings of \cite{Birse:2010jr}. The explanation can also be that the lower cutoffs neglect relevant loop contributions from the iterated one-pion exchange at LO, reflected in the large error observed in the $\epsilon_1$ phase shift for $\Lambda=500$~MeV in \cref{fig:phase_3S1}---an effect that can be challenging to correct perturbatively. Detailed studies of the convergence of the perturbative expansion and possible connections to the $c_i$ LECs are left for future studies. It could also be interesting to address how the convergence is affected by using different regularization schemes, e.g., spectral function regularization \cite{Epelbaum:2003xx,Epelbaum:2003gr}, and by including the $\Delta(1232)$ isobar as an explicit degree of freedom. 

\section{Conclusions}\label{sec:conclusions}
In this study, we have computed LETs for \np{} scattering in \chiEFT{} to investigate the low-energy (long-range) behavior of the Long and Yang PC. The following conclusions can be made:

\begin{enumerate}[(i)]
    \item LETs can be accurately computed in perturbation theory; both by directly using the scattering amplitudes (\cref{eq:F_T_0,eq:F_T_1,eq:F_T_2,eq:F_T_3}) and by using phase shifts (\cref{eq:F_phase}). 

    \item It is important to consider CIB in the one-pion exchange potential due to the pion mass splitting in the $^1S_0$ partial wave to simultaneously describe empirical phase shifts and ERE parameters. Without considering CIB, we show that the phase shifts converge well at \NNNLO{} up to $\Tl = 100$~MeV, while LETs do not improve beyond NLO and a significant tension with empirical ERE parameters remains at \NNNLO. When considering CIB, however, the tension in the LETs at \NNNLO{} essentially disappears while the phase shifts still show a good convergence up to $\Tl = 100$~MeV.

    \item In the $^3S_1$ partial wave we show that CIB does not need to be considered since it has a negligible effect on both phase shifts and LETs. The LETs reproduce the empirical ERE parameters while an accurate description of phase shifts up to (at least) $\Tl=100$~MeV is kept---as long as the cutoff is taken large enough ($\Lambda \gtrsim 750$~MeV). For smaller cutoffs, a large residual cutoff dependence in both phase shifts and LETs is observed at \NNNLO{}. Further studies are needed to assess the origin of this larger-than-expected residual dependence.

    \item The observed simultaneous accuracy of LETs and low-energy phase shifts shows that the low-energy description of the nuclear force is captured in the Long and Yang PC for the considered $S$-waves. The convergence pattern of the phase shifts in both $S$-waves shows that sub-leading interactions, in particular the two-pion exchanges, are amendable to a perturbative treatment at least up to $\Tl=100$~MeV.
\end{enumerate}
In this study, we did not consider adding the CIB due to the pion mass splitting perturbatively, something that can be investigated further. The construction of an improved description at LO in the $^1S_0$ partial wave should also be explored. This can be achieved, e.g., by promoting potential contributions that better capture the effective range, see for example~Refs.~\cite{Long:2013cya,Epelbaum:2015sha}. 

It would also be valuable to quantify the extent to which neglected sources of uncertainty in both data and models, along with choices made during the LEC calibration, influence the results. This can be effectively achieved within a Bayesian framework \cite{Furnstahl:2014xsa}, wherein LECs are treated as stochastic variables. However, conducting a Bayesian inference of LECs may encounter challenges, including the risk of overfitting, and the emergence of spurious correlations among LECs at different orders. The latter is particularly prominent in a perturbative PC where the perturbative corrections to LECs naturally exhibit strong correlations. 
The results of this study confirm the effectiveness of LETs also in this perturbative PC which can be used as a tool to counteract overfitting LECs to high-energy data and ensuring that the fidelity of the low-energy description is maintained.

\backmatter

\bmhead{Acknowledgments}
O.T is grateful to Andreas Ekström, Christian Forssén and Daniel Phillips for helpful discussions as well as for providing feedback on the manuscript. O.T also acknowledges insightful comments from the anonymous referee about the CIB effect in the $^1S_0$ channel. This work was supported by the European Research Council (ERC) under the European Unions Horizon 2020 research and innovation program (Grant Agreement No.~758027) and the Swedish Research Council (Grant No.~2020-05127). 

\begin{appendices}

\newpage
\section{Additional tables\label{app:tables}}
This appendix contains some additional tables that summarize some LET results from the literature for the $^1S_0$ partial wave (\cref{tab:ERE_1S0_ref}) and the $^3S_1$ partial wave (\cref{tab:ERE_3S1_ref}).
\begin{table}[h]
    \caption{Collection of LET results in the $^1S_0$ partial wave from several studies. The parameters marked with a star ($*$) were used in the inference in the respective studies.}
\begin{center}

\begin{tabular}{l|ccccc}
    \toprule
    $^1S_0$ partial wave & $a$ [fm] & $r$ [fm] & $v_2$ [fm$^3$]& $v_3$ [fm$^5$] & $v_4$ [fm$^7$]\\
    \midrule
    Mean Ref.~\cite{NavarroPerez:2014ovp} & $-23.735(16)$ & 2.68(3) & $-0.48(2)$ & 3.9(1) & $-19.6(5)$ \\
      \midrule 
    \NNLO\footnotemark[1] WPC (DR) Ref.~\cite{Epelbaum:2003xx} & $-23.936$ & 2.73 & $-0.46$ & 3.8 & $-19.1$ \\
    \midrule
    NLO KSW from Ref.~\cite{Cohen:1998jr} & $*$ & $*$ & $-3.3$ & 18 & $-108$ \\
    \midrule
    LO  from Ref.~\cite{Epelbaum:2012ua} & $*$ & 1.5 & $-1.9$ & 9.6(8) & $-37(10)$ \\
    NLO (pert.) from Ref.~\cite{Epelbaum:2015sha} & $*$ & $*$ & $-0.54(3)$& 4.6(1)& $-29.3(5)$ \\

    \botrule
\end{tabular}
\end{center}
\footnotetext[1]{Note that the chiral orders are referred to differently in Ref.~\cite{Epelbaum:2003xx}. The orders NLO and \NNLO{} in Ref.~\cite{Epelbaum:2003xx} correspond to \NNLO{} and \NNNLO{} in this work.}

\label{tab:ERE_1S0_ref}
\end{table}

\begin{table}[h]
    \caption{Collection of LET results in the $^3S_1$ partial wave. The parameters marked with a star ($*$) were used in the inference in the respective studies.}
\begin{center}
\begin{tabular}{c|ccccc}
    \toprule
    $^3S_1$ partial wave & $a$ [fm] & $r$ [fm] & $v_2$ [fm$^3$]& $v_3$ [fm$^5$] & $v_4$ [fm$^7$]\\
    \midrule
    NijmII (Ref.~\cite{PavonValderrama:2005ku}) & 5.419 & 1.753 & 0.0453 & 0.658 & -4.191 \\
    Reid93 (Ref.~\cite{PavonValderrama:2005ku}) & 5.423 & 1.756 & 0.0327 & 0.658 & -4.193 \\
    NijmII (Ref.~\cite{NavarroPerez:2014ovp}) & 5.4197(3) & 1.75343(3) & 0.04545(1) & 0.6735(1) & -3.9414(8) \\
    \midrule
    LO WPC (Ref.~\cite{Epelbaum:2012ua}) & $*$ & 1.60 & $-0.05$ & 0.8(1) & $-4(1)$ \\
    \NNLO\footnotemark[1] WPC&&&&& \\ 
     (DR) (Ref.~\cite{Epelbaum:2003xx}) & 5.416 & 1.756 & 0.04 & 0.67 & $-4.1$ \\
    \midrule
    NLO KSW (Ref. \cite{Cohen:1998jr})& $*$ & $*$ & -0.95 & 4.6 & -25.0 \\
    \botrule
\end{tabular}
\end{center}
\footnotetext[1]{Note that the chiral orders are referred to differently in Ref.~\cite{Epelbaum:2003xx}. The order \NNLO{} in Ref.~\cite{Epelbaum:2003xx} corresponds to \NNNLO{} in this work.}
\label{tab:ERE_3S1_ref}
\end{table}

\section{Computing phase shifts perturbatively\label{app:pert_phase}}
The $^1S_0$ phase shift, $\delta(\pon)$, and on-shell scattering amplitude, $T(k,k)$, are related through \cref{eq:ST_pw} which simplifies to 
\begin{equation}
    e^{2i\delta} = 1- i\pi \pon \mN  T(k,k),
    \label{eq:S_l0}
\end{equation}
where the quantum numbers are suppressed from the notation.
By expressing both the phase shift and the amplitude in contributions at each chiral order according to \cref{eq:T-sum,eq:phase-sum} and expanding both sides of \cref{eq:S_l0} the following relations are obtained
\begin{align}
    \exp\left(2i\delta^{(0)}\right) &= 1- i\pi \pon \mN  T^{(0)}(k,k) \\
    2i\delta^{(1)} &= -i\pi \pon \mN T^{(1)}(k,k)  \exp\left(-2i\delta^{(0)}\right) \\
    2i\delta^{(2)} - 2 \left(\delta^{(1)}\right)^2 &=  -i\pi \pon \mN T^{(2)}(k,k)\exp\left(-2i\delta^{(0)}\right)\\
     2i\delta^{(3)} - 4 \delta^{(1)}\delta^{(2)} - \frac{4i}{3}\left(\delta^{(1)}\right)^3 &=-i\pi \pon \mN T^{(3)}(k,k) \exp\left(-2i\delta^{(0)}\right).
\end{align}
From these equations, the phase shift corrections, $\delta^{(\nu)}$, can be obtained.

The computation in the coupled $\tS$ channel is slightly more involved. The BB parametrization \cite{Blatt:1952zz} of the unitary $2\times 2$ $S$-matrix reads
\begin{equation}
    S = \begin{pmatrix}
        \cos{\epsilon} &  \  -\sin{\epsilon} \\
        \sin{\epsilon} &  \ \cos{\epsilon}
    \end{pmatrix}\begin{pmatrix}
        e^{2i\delta_-} & \ 0 \\
        0 & \ e^{2i\delta_+}
    \end{pmatrix} \begin{pmatrix}
        \cos{\epsilon} & \ \sin{\epsilon} \\
        -\sin{\epsilon} & \ \cos{\epsilon}
    \end{pmatrix}
\end{equation}
where $\delta_-$ and $\delta_+$ are the eigen phase shifts corresponding to the $^3S_1$ and $^3D_1$ partial waves, respectively and $\epsilon$ is the mixing angle. Using \cref{eq:ST_pw} the phase shifts can be expressed as
\begin{align}
    \epsilon &= \frac{1}{2}\arctan\left(\frac{2T^{sj}_{-+}}{T^{sj}_{--} - T^{sj}_{++}}\right) \label{eq:eps}\\
    \delta_-\left(\pon\right) &= -\frac{i}{2} \log\left[1- \pi i \mN \pon\left( \frac{T^{sj}_{--}+T^{js}_{++}}{2} + \frac{T^{sj}_{-+}}{\sin{2\epsilon}}\right)\right] \label{eq:delta_m}\\
    \delta_+\left(\pon\right) &= -\frac{i}{2} \log\left[1- \pi i \mN \pon\left( \frac{T^{sj}_{--}+T^{sj}_{++}}{2} - \frac{T^{sj}_{-+}}{\sin{2\epsilon}}\right)\right], \label{eq:delta_p}
\end{align}
where the $\pm$ notation represent $\l',\ \l = j\pm 1$. From the computed on-shell amplitudes $T^{(\nu)sj}_{\l'\l}$ for orders $\nu=0,1,2,3$ the corresponding phase shifts $\big(\delta^{(\nu)}_-,\delta^{(\nu)}_+,\epsilon^{(\nu)}\big)$, $\nu=0,1,2,3$ are obtained by Taylor expanding \cref{eq:eps,eq:delta_m,eq:delta_p} and matching chiral orders. This is completely analogous to the treatment of the $^1S_0$ partial wave and the treatment of the Stapp parametrization \cite{Stapp:1956mz} in Refs.~\cite{PhysRevC.85.034002,Thim:2024yks}.
\end{appendices}

\bibliography{sn-bibliography}% common bib file

\end{document}